# Radiative Balance Between Photon Emission From Surface And In Volume

Tom Markvart

Engineering Materials, University of Southampton, Southampton SO17 1BJ, UK, and
Centre for Advanced Photovoltaics, Czech Technical University, 166 36 Prague 6, Czech Republic

Abstract

An expression is derived which links the probability of photon reabsorption (recycling) in a material with other optical parameters. We discuss the application of this relation in several typical situations which are relevant in the operation of devices such as solar cells and light-emitting diodes.

The study of balance between the absorption and emission of radiation in matter has a long and distinguished history. Beginning with Kirchhoff's law,[1] it provided the background for Planck in his derivation of the black body radiation formula leading to quantum physics,[2] and paved the way for Einstein's discovery of stimulated emission and invention of the laser.[3] More recently, different facets of the balance appear in the use of light trapping to enhance the output of both solar cells[4,5] and light-emitting diodes,[6] and in the suggestions that photon recycling (the re-absorption of internally re-emitted light) may enhance the operation of solar cells[7,8,9] (Fig. 1). This paper derives a simple relation which, by providing a link between the probability of photon recycling and other optical parameters, can be used to elucidate and explain the role of photon recycling in modern optoelectronic devices.

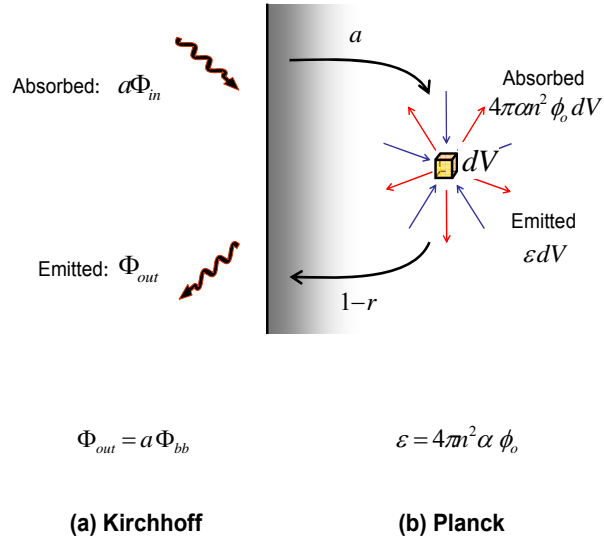

$$\Phi_{out} = a \Phi_{bb} \qquad \varepsilon = 4\pi n^2 \alpha \, \phi_o$$

**(a) Kirchhoff**　　　　**(b) Planck**

Fig. 1. The Kirchhoff (surface) and Planck (volume) expressions for radiative balance in an absorbing medium (adapted from ref. 12) which are linked in this paper via the probability of photon reabsorption / recycling $r$.

Assuming, for simplicity, that the emitting medium is surrounded by vacuum, the monochromatic photon flux emitted per unit time by a surface is given by

$$\Phi_S = e\Phi_{bb} = e\mathcal{E}\phi_o \qquad (1)$$

where $e$ is Kirchhoff's emissivity,[8] $\Phi_{bb}$ is the black-body emission rate, $\phi_o$ is the black-body photon flux per unit area per unit solid angle in vacuum, and $\mathcal{E}$ is the étendue of the emitted beam (see, for example, ref. 10).

The rate of internal photon emission is

$$\Phi_V = \varepsilon V = 4\pi n^2 \alpha \phi_o V \qquad (2)$$

where $\varepsilon$ is the photon (rather than energy) analogue of Planck's emission coefficient (rate of photon emission per unit volume), $\alpha$ is the absorption coefficient and $n$ is the refractive index of the medium. The second equality follows from the detailed balance between absorption and emission in the volume as formulated by Planck.[2]

Since all internally emitted photons will leave the medium unless removed by reabsorption (with average probability $r$) we can write

$$\Phi_S = (1-r)\Phi_V \qquad (3)$$

and substituting from (1) and (2),

$$1 - r = \frac{a}{\ell_{opt}\alpha} \qquad (4)$$

where Kirchhoff's law was used to replace emissivity $e$ by absorptivity $a$, and we introduced a quantity $\ell_{opt}$ with the dimension of length:

$$\ell_{opt} = \frac{4\pi n^2 V}{\mathcal{E}} \qquad (5)$$

For a planar structure of thickness $d$, $\ell_{opt}$ reduces to $4n^2 d$. In the limit of high absorptivity (for example, a black body), Eq. (4) gives an upper bound on the reabsorption probability in the form

$$r \approx 1 - \frac{1}{\ell_{opt}\alpha} \qquad (6)$$

Equation (4) will find a useful application to light emitting diodes where $1-r$ has the meaning of photon escape probability $\eta_{esc}$. Relation (4) therefore provides a convenient expression for $\eta_{esc}$ in terms of the optical parameters and dimensions of the device. In solar cells or LEDs where light capture or emission is enhanced by light trapping, $\ell_{opt}$ can be interpreted as the mean path for photons absorption.[11] Indeed, in a model where light trapping is described in terms of scattering, the probability of absorption is independent of whether light is incident from the exterior or interior of the layer, and we can therefore set $a = r$. Equation (4) then gives

$$a = 1 - r = \frac{\ell_{opt}\alpha}{1 + \ell_{opt}\alpha}$$

providing a more general background for the result originally suggested by Tiedje et al.[5] (see also ref. 12). The reabsorption probabilities for a layer with or without light trapping are compared in Fig. 2.

Acknowledgement

This work was supported by the UK EPSRC SUPERGEN program "Photovoltaic materials for the 21st century" EP/F029624/2.

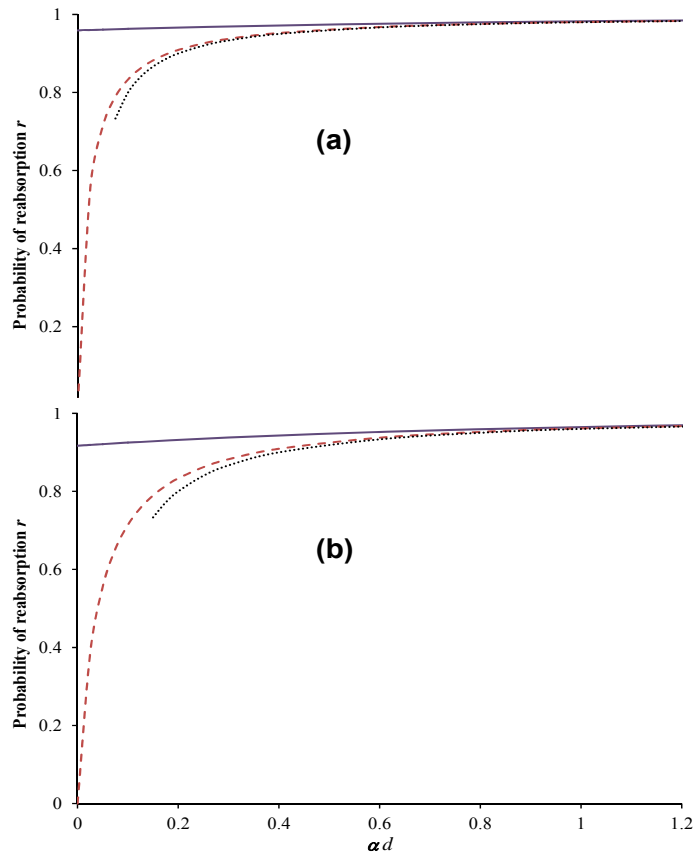

Fig. 2. The probability of reabsorption $r$ as a function of the optical density $\alpha d$ for (a) $n^2 = 50$ (typical for inorganic semiconductors such as silicon or gallium arsenide) and (b) $n^2 = 25$, typical for e.g. perovskites. Full lines: flat surface; dashed lines: light trapping structures; dotted lines: the black body limit (6).